\documentclass[trackchanges,resetfootnote]{aastex701}

\usepackage{wrapfig}
\usepackage{color}
\usepackage{multirow}

\newcommand\mps{m s$^{-1}$~}
\newcommand\mpsp{m s$^{-1}$}
\newcommand\dets{\emph{dET}~}
\newcommand\detp{\emph{dET}}

\newcommand{\PSUAA}{Department of Astronomy and Astrophysics, 525 Davey Laboratory, 251 Pollock Road, Penn State, University Park, PA, 16802, USA}
\newcommand{\PSUCEHW}{Center for Exoplanets and Habitable Worlds, 525 Davey Laboratory, 251 Pollock Road, Penn State, University Park, PA, 16802, USA}
\newcommand{\PSUARC}{Astrobiology Research Center, 525 Davey Laboratory, 251 Pollock Road, Penn State, University Park, PA, 16802, USA}

\newcommand{\UA}{Steward Observatory, University of Arizona, 933 N.\ Cherry Ave, Tucson, AZ 85721, USA}

\newcommand{\GSFC}{NASA Goddard Space Flight Center, Greenbelt, MD 20771, USA}
\newcommand{\NOAO}{U.S. National Science Foundation National Optical-Infrared Astronomy Research Laboratory, 950 N.\ Cherry Ave., Tucson, AZ 85719, USA}

\newcommand{\Macquarie}{School of Mathematical and Physical Sciences, Macquarie University, Balaclava Road, North Ryde, NSW 2109, Australia}

\newcommand{\CUBoulderEE}{Electrical, Computer \& Energy Engineering, 440 UCB, University of Colorado, Boulder, CO 80309, USA}
\newcommand{\CUBoulder}{Department of Physics, 390 UCB, University of Colorado, Boulder, CO 80309, USA}
\newcommand{\JPL}{Jet Propulsion Laboratory, California Institute of Technology, 4800 Oak Grove Drive, Pasadena, California 91109}

\newcommand{\UCI}{Department of Physics \& Astronomy, The University of California, Irvine, Irvine, CA 92697, USA}
\newcommand{\Carleton}{Carleton College, One North College St., Northfield, MN 55057, USA}
\newcommand{\Carnegie}{Earth and Planets Laboratory, Carnegie Science, 5241 Broad Branch Road, NW, Washington, DC 20015, USA}

\newcommand{\TIFR}{Department of Astronomy and Astrophysics, Tata Institute of Fundamental Research, Homi Bhabha Road, Colaba, Mumbai 400005, India}

\newcommand{\Amherst}{Department of Physics and Astronomy, Amherst College, 25 East Drive, Amherst, MA 01002, USA}

\newcommand{\Schmidt}{Astrophysics \& Space Institute, Schmidt Sciences, New York, NY 10011, USA}
\begin{document}

\title{A Revised Mass and Period for the Habitable Zone super-Earth GJ~3378~b: A Planet Straddling the Cosmic Shoreline \footnote{Based on data obtained with the HET/HPF and WIYN/NEID instruments}}

\author[0000-0003-0149-9678]{Paul Robertson}
\affiliation{\UCI}
\email[show]{paul.robertson@uci.edu}

\author[0000-0002-7714-6310]{Michael Endl}
\affiliation{Center for Planetary Systems Habitability, The University of Texas, Austin TX 78712 USA}
\affiliation{Department of Astronomy and McDonald Observatory, The University of Texas, Austin TX 78712 USA.}
\email{mike@astro.as.utexas.edu}  

\author[0000-0001-9662-3496]{William D. Cochran}
\affiliation{Center for Planetary Systems Habitability, The University of Texas, Austin TX 78712 USA}
\affiliation{Department of Astronomy and McDonald Observatory, The University of Texas, Austin TX 78712 USA.}
\email{wdc@astro.as.utexas.edu}

\author[0000-0001-7409-5688]{Gudmundur Stefansson}
\affiliation{\Schmidt}
\affil{Anton Pannekoek Institute for Astronomy, University of Amsterdam, Science Park 904, 1098 XH Amsterdam, The Netherlands}
\email{gstefansson@schmidtsciences.org}

\author[0000-0001-9596-7983]{Suvrath Mahadevan}
\affiliation{\PSUAA}
\affiliation{\PSUCEHW}
\affiliation{\PSUARC}   
\email{suvrath@astro.psu.edu}

\author[0000-0003-4835-0619]{Caleb I. Ca\~nas}
\affiliation{Southeastern Universities Research Association, Washington, DC 20005, USA}
\affiliation{\GSFC}
\email{c.canas@nasa.gov}

\author[0009-0007-6159-2520]{Gogod James}
\affiliation{\UCI}
\email{gtjames@uci.edu}

\author[0009-0001-6669-264X]{Roan Arendtsz}
\affiliation{\Carleton}
\email{arendtszr@carleton.edu}

\author[0000-0002-4788-8858]{Ryan C. Terrien}
\affiliation{\Carleton}
\email{rterrien@carleton.edu}

\author[0000-0003-4384-7220]{Chad F.\ Bender}
\affiliation{\UA}
\email{cbender@arizona.edu}

\author[0000-0002-2144-0764]{Scott A.\ Diddams}
\affiliation{\CUBoulderEE}
\affiliation{\CUBoulder}
\email{scott.diddams@colorado.edu}

\author[0000-0002-0078-5288]{Mark R.~Giovinazzi}
\affiliation{\Amherst}
\email{mgiovinazzi@amherst.edu}

\author[0000-0002-5463-9980]{Arvind F.\ Gupta}
\affiliation{\NOAO}
\email{arvind.gupta@noirlab.edu}

\author[0000-0003-1312-9391]{Samuel Halverson}
\affiliation{\JPL}
\email{samuel.halverson@jpl.nasa.gov}

\author[0000-0001-8401-4300]{Shubham Kanodia}
\affiliation{\Carnegie}
\email{skanodia@carnegiescience.edu}

\author[0000-0001-9626-0613]{Daniel M.\ Krolikowski}
\affiliation{\UA}
\email{krolikowski@arizona.edu}

\author[0000-0002-9632-9382]{Sarah E.\ Logsdon}
\affiliation{\NOAO}
\email{sarah.logsdon@noirlab.edu}

\author[0000-0001-8720-5612]{Joe P.\ Ninan}
\affiliation{\TIFR}
\email{indiajoe@gmail.com}

\author[0009-0005-5520-1648]{Claire J. Rogers}
\affiliation{\UCI}
\email{c.j.rogers@uci.edu}

\author[0000-0001-8127-5775]{Arpita Roy}
\affiliation{\Schmidt}
\email{arpita308@gmail.com}

\author[0000-0002-4046-987X]{Christian Schwab}
\affiliation{\Macquarie}
\email{mail.chris.schwab@gmail.com}

\begin{abstract}

The nearby ($d = 7.7$ pc) M4V star GJ~3378 is a target of our radial velocity (RV) exoplanet survey of fully convective stars in the Solar neighborhood with the near-infrared spectrograph HPF on the Hobby-Eberly Telescope (HET) at McDonald Observatory.  Recently, Moutou et al.~(2024) announced the discovery of an $m\sin i = 5.26^{+0.94}_{-0.97} M_\oplus$ planet, GJ 3378b, with an orbital period of $24.73 \pm 0.06$ days, based on SPIRou RV data.  Here, we present our HPF RVs for GJ 3378, as well as  additional Doppler spectroscopy from the extreme precision NEID Spectrometer on the WIYN telescope at Kitt Peak National Observatory. We have analyzed the HPF+NEID RVs jointly with the published RVs from the CARMENES and SPIRou spectrometers. We present an orbital model for GJ 3378b that differs significantly from the Moutou et al.~solution.  The joint RV model reduces the orbital period to $P = 21.45 \pm 0.01$d and the minimum mass to $m \sin i = 2.3 \pm 0.4 M_\oplus$.  The shortened orbital distance remains within the conservative circumstellar liquid-water habitable zone (HZ), while the reduced mass increases the likelihood that the planet has a terrestrial composition.  The revised planet properties place it near the ``cosmic shoreline," where planets in the HZs of M dwarfs may lose their atmospheres due to radiative stripping.  

\end{abstract}


\section{Introduction} 

Direct detection of temperate, terrestrial exoplanets is a key science case of next-generation telescopic facilities both on the ground \citep{johns2012,quanz2015,skidmore2015} and in space \citep{gaudi2020,luvoir2019,quanz2022,hwonas2023}.  A specific advantage of ground-based 30\,m-class telescopes is the ability to image Earthlike planets in the Habitable Zones \citep[HZs;][]{kopparapu2013} of nearby M stars.  Owing to the minuscule intrinsic luminosities of M dwarfs ($L_\star < 0.1 L_\odot$), their HZ planets reside at angular separations that are only accessible to the largest apertures, or at long interferometeric baselines.  Even for 30\,m-class telescopes, targets must typically lie within 10 pc of the Solar System to make direct imaging feasible.  It is therefore of critical importance that we intensively survey the nearest M dwarfs with radial velocities (RVs) to identify small HZ planets that will be ideal targets for these upcoming facilities.  It is likewise vital that we place meaningful upper limits on systems where no HZ planets are detected, and identify planets which may have dynamical implications for the existence or long-term orbital evolution of planets within the HZ \citep{kane2024,kaneburt2024}.

RVs are the most effective technique for detecting HZ exoplanets around the nearest M stars.  For the typical HZ extents of M stars, the geometric transit probability $R_\star/a$ is of order 1\%; with $<300$ M dwarfs within 10 pc \citep{henry2018}, the transit method is not likely to discover many targets accessible to direct imaging.  Likewise, terrestrial HZ exoplanets around M dwarfs induce sub-microarcsecond astrometric motions on their host stars, which is highly challenging to recover with the precision and observing cadence of Gaia \citep{gaia2016,gaia2024}.

Fortunately, a fleet of new Doppler spectrometers in the optical \citep{quirrenbach2014,jurgenson2016,gibson2016,seifahrt2018,schwab2019,pepe2021} and infrared \citep{tamura2012,mahadevan2014,donati2020,malo2024} are achieving sensitivity to terrestrial-mass exoplanets orbiting nearby M stars.  The spectra of cool M stars contain an immense amount of Doppler information content in both the red-optical and the near-infrared \citep{reiners2020}, and the relatively large planet-to-star mass ratios for M dwarfs make terrestrial planets accessible at or near the 1 \mps RV precision level.  Indeed, the recent RV discoveries of sub-Earth-mass exoplanets orbiting the M dwarfs Proxima Centauri \citep{faria2022} and Barnard's Star \citep{gonzalezhernandez2024,basant2025} represent the lowest-amplitude exoplanet discoveries with the Doppler technique to date.

We are conducting a blind Doppler survey of exoplanets using the near-infrared (NIR) Habitable-zone Planet Finder Spectrometer \citep[HPF;][]{mahadevan2014} on the 10\,m Hobby-Eberly Telescope \citep[HET;][]{ramsey1998,hill2021} at McDonald Observatory in Texas.  Our survey is monitoring all accessible stars of spectral type M3 and later within 8 pc (36 stars), and a selection of 40 more distant mid-to-late M dwarfs for exoplanets.  Our survey strategy is designed to leverage HPF's outstanding $\sim 1$ \mps NIR RV stability \citep{metcalf2019,lubin2021} to detect terrestrial-mass exoplanets in the HZs of our targets.

In this manuscript, we present a confirmation, and significant revision, of the HZ exoplanet orbiting the nearby \citep[$d = 7.7$ pc;][]{bailerjones2021} M4 dwarf GJ 3378.  GJ 3378 exhibits a periodic RV signal at just over 20 days.  It was originally identified by the CARMENES exoplanet survey \citep{sabotta2021} with an ``unsolved" designation.  \citet{moutou2024} announced the signal as an exoplanet candidate using RVs from the infrared SPIRou Spectrometer.  Moutou et al.~described the planet as having an orbital period of 24.73 days, a modest eccentricity of $e = 0.36^{+0.13}_{-0.16}$, and a Doppler amplitude $K = 3.08^{+0.60}_{-0.58}$ \mpsp, corresponding to a minimum mass $m \sin i = 5.26^{+0.94}_{-0.97} M_\oplus$.  This orbital period implies an orbital separation $a = 0.106 \pm 0.003$ AU, which is within the conservative HZ according to \citet{kopparapu2013}, albeit at a mass range that likely excludes Earthlike compositions \citep[e.g.][]{parc2024}.

\begin{figure}[]
    \centering
    \includegraphics[width=\textwidth]{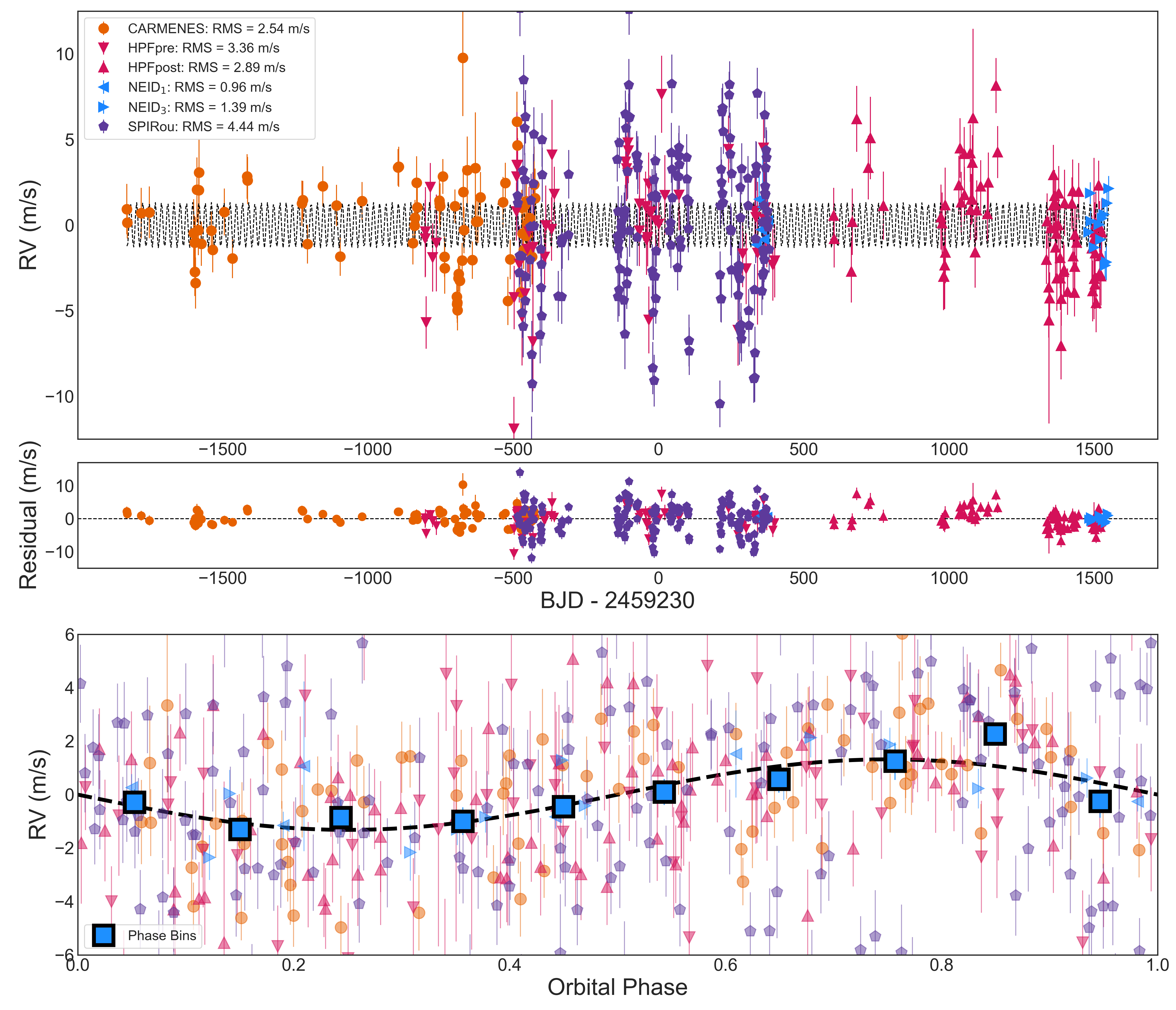}
    \caption{Radial velocity data for GJ 3378.  \emph{Top}: the combined RV time series.  The orbit model is shown as a dashed line.  \emph{Middle}: RV residuals to a 1-planet model.  \emph{Bottom}: RVs phase-folded to the period of the exoplanet.  Phase-binned RVs are shown as squares.}
    \label{fig:orbit}
\end{figure}

Here, we add 137 precise RVs from HPF to the previously-published velocities, as well as 18 from the ultra-stable NEID spectrometer \citep{schwab2019} on the 3.5\,m WIYN Telescope\footnote{The WIYN Observatory is a joint facility of the NSF's National Optical-Infrared Astronomy Research Laboratory, Indiana University, the University of Wisconsin-Madison, Pennsylvania State University, and Princeton University.} at Kitt Peak National Observatory (KPNO).  These new observations confirm the previously-identified planet candidate, but we find a significantly different orbital solution.  Specifically, our model indicates a shorter, circular orbit, with a period of $21.45 \pm 0.01$d, and a much smaller RV amplitude $K = 1.3 \pm 0.2$ \mpsp.  The revised orbit means the planet remains within the conservative HZ, but its minimum mass decreases to $m \sin i = 2.3 \pm 0.4 M_\oplus$, potentially consistent with a terrestrial composition.

\section{Data}

\subsection{HPF Spectroscopy}

HPF is a near-infrared ($\lambda \lambda =810-1280$ nm), highly stabilized \citep{stefansson2016}, fiber-fed \citep{halverson2015,kanodia2021} spectrometer on the 10\,m Hobby-Eberly Telescope \citep{ramsey1998,hill2021}.  GJ 3378 is a target of our ongoing survey of nearby fully-convective M dwarfs, and we have observed it 137 times between November 2018 and March 2025.  Our observations were acquired through HET's queue-scheduled system \citep{shetrone2007}, and each visit to the target typically consisted of two up-the-ramp exposures of 969s each.

Spectral extraction was performed using the HPF H2RG pipeline \citep{kaplan2019,ninan2019}.  This pipeline produces 1D extracted spectra, from which we estimate RVs using a version of the \texttt{SERVAL} template-matching algorithm \citep{zechmeister2018} adapted for use with HPF data \citep{stefansson2020,stefansson2023}.  

The \texttt{HPF-SERVAL} implementation also provides spectroscopic activity indicators such as the NIR calcium triplet line strengths, and line-shape metrics such as dLW.  An additional especially valuable indicator within the HPF bandpass is the 12435.67 \AA~neutral potassium (KI) equivalent width (EW), which is not computed by \texttt{HPF-SERVAL}.  The KI EW is sensitive to variability driven by the Zeeman effect, and has been shown to be effective at revealing rotation signals from old, slowly-rotating M dwarfs \citep{terrien2022}.  The KI EW measurements are corrected for slow drifts in the continuum level, as described in Arendtsz et al.~(2026, submitted).  We have excluded from our analysis two KI EW values: one from 2021 December 22 which had very low signal to noise, and one $\sim10\sigma$ outlier from 2025 February 15.

For the analysis presented here, we have binned all HPF RVs and associated activity metrics collected within a single 2-exposure visit.  The HPF vacuum vessel was opened in May 2022 for instrumental maintenance, causing a break in the instrument's RV zero point.  We treat observations before and after this break as separate time series in order to model the separate zero-point offsets as free parameters.  The two sets of RVs---referred to hereafter as HPF$_{\textrm{pre}}$ and HPF$_{\textrm{post}}$---include 57 and 80 binned RVs, respectively.  HPF$_{\textrm{pre}}$ RVs have an RMS scatter of 3.36 \mps and a median uncertainty of $2.04$ \mpsp.  The statistics for HPF$_{\textrm{post}}$ are somewhat improved, with an RMS of 2.89 \mps and median uncertainty of 1.73 \mpsp.  The combined time baseline for both RV sets spans 2328 days.

\subsection{NEID Spectroscopy}

NEID \citep{schwab2019} is a broadband-optical ($\lambda \lambda = 380-930$ nm), highly stabilized \citep{robertson2019}, fiber-fed \citep{kanodia2023} Doppler spectrometer on the 3.5\,m WIYN Telescope at Kitt Peak National Observatory in Arizona.  NEID is being used to conduct the NEID Earth Twin Survey \citep[NETS;][]{gupta2021,gupta2025}, a long-baseline, blind RV search for small exoplanets orbiting nearby stars.  GJ 3378 was observed a handful of times in the 2021B observing semester by NETS as part of a broader effort to determine NEID's achievable RV precision on potential survey targets.  More recently, we conducted a high-cadence campaign in the 2025A semester with NEID on GJ 3378 in order to confirm or rule out the planet candidate described herein.

Our NEID observations of GJ 3378 use the spectrometer in the high-resolution (HR) mode, which offers a resolving power $R \sim 110\,000$.  All observations include two 900s exposures per visit, with one (2022 Jan 14) including a third, truncated exposure of 137s.  Basic data reduction and 1D spectral extraction were performed automatically using version 1.4.1 of the standard NEID Data Reduction Pipeline\footnote{\url{https://neid.ipac.caltech.edu/docs/NEID-DRP}} \citep[DRP;][]{bender2022}.  The DRP offers RV extraction as well, but the RVs are estimated using the binary mask technique \citep[e.g.][]{baranne1996}, which is suboptimal for very cool stars with blended lines.  Thus, rather than using the DRP RVs, we have measured RVs from the DRP-extracted spectra using an implementation of \texttt{SERVAL} adapted for use with NEID \citep{stefansson2022}.  For our analysis, we have binned all NEID data taken within a single night.

NEID experienced multiple RV zero point offsets between our 2021B and 2025A observations.  According to the definitions adopted by \citet{gupta2025}, our 2021B RVs come from Science Run 1, and the 2025A observations are from Science Run 3.\footnote{See more details at \url{https://neid.ipac.caltech.edu/docs/NEID-DRP/rveras.html}}  Thus, as with HPF, we have separated the NEID RVs into two separate time series, dubbed NEID$_1$ and  NEID$_3$.  NEID$_1$ includes 6 visits spanning 40 days, and has an RMS of $0.96$ \mps with a median uncertainty of $0.51$ \mpsp.  NEID$_3$ contains 12 RVs across 69 days, exhibits an RMS of $1.39$ \mps, and has a median uncertainty of $0.74$ \mpsp.

\subsection{Archival CARMENES Spectroscopy}

GJ 3378 has been monitored by the stabilized, dual-arm optical/NIR CARMENES spectrometer \citep{quirrenbach2014} on the 3.5\,m telescope at Calar Alto Observatory in Almer\'ia, Spain.  80 observations of GJ 3378 were made public via Data Release 1 of the CARMENES Guaranteed-Time Observations \citep{ribas2023}.  These RVs are from the visible (VIS) channel of CARMENES, which spans wavelengths from 520-960 nm.  The data release provided extracted radial velocities, as well as a host of spectral activity indicators such as activity-sensitive line strengths (e.g.~H$\alpha$ and the NIR calcium triplet) and line-shape parameters such as the differential line width \citep[dLW;][]{zechmeister2018}.  The data provided come from the optical arm of the CARMENES spectrometer.

Here, we use the RVs extracted using the \texttt{SERVAL} template-matching algorithm \citep{zechmeister2018}.  We restrict our analysis to velocities corrected with CARMENES nightly zero points (NZPs; designated \texttt{AVC} in the CARMENES data release), and remove any observations with data quality flags.  The resulting time series includes 78 RVs, which span a temporal baseline of 1404 days.  The RVs exhibit an RMS scatter of 2.54 \mps and a median uncertainty of 1.26 \mpsp.

\begin{table}[]
    \centering
    \begin{tabular}{l c c}
        \hline \hline
        Parameter & Value & Reference \\
        \hline
        \\
        \multicolumn{3}{c}{\it Measured Stellar Properties} \\
        \\
        Right Ascension $\alpha$ (ICRS) & 06h01m11.046s & \citet{gaiadr3} \\
        Declination $\delta$ (ICRS) & +59$^\circ$35$^\prime$49.884$\arcsec$ & \citet{gaiadr3} \\
        Parallax (milliarcseconds) & $129.30 \pm 0.03$ & \citet{gaiadr3} \\
        RA proper motion $\mu_\alpha \cos \delta$ (mas yr$^{-1}$) & $-104.94 \pm 0.02$ & \citet{gaiadr3} \\
        Dec proper motion $\mu_\delta$ (mas yr$^{-1}$) & $-928.63 \pm 0.03$ & \citet{gaiadr3} \\
        $G$ magnitude & $10.417 \pm 0.003$ & \citet{gaiadr3} \\
        $J$ magnitude & $7.47 \pm 0.02$ & \citet{monet2003} \\
        $H$ magnitude & $6.95 \pm 0.03$ & \citet{monet2003} \\
        $K$ magnitude & $6.64 \pm 0.02$ & \citet{monet2003} \\
        $W1$ magnitude & $6.53 \pm 0.04$ & \citet{wise} \\
        $W2$ magnitude & $6.31 \pm 0.02$ & \citet{wise} \\
        $W3$ magnitude & $6.22 \pm 0.02$ & \citet{wise} \\
        $W4$ magnitude & $6.13 \pm 0.05$ & \citet{wise} \\
        \\
        \multicolumn{3}{c}{\it Derived Stellar Properties}
        \\
        Distance $d$ (pc) & $7.730^{+0.002}_{-0.001}$ & \citet{bailerjones2021} \\
        Effective Temperature $T_{eff}$ (K) & {$3340 \pm 60$} & \cite{mann2015} \\
        Surface Gravity $\log g_\star$ (cgs) & {$4.97 \pm 0.06$} & \cite{mann2015} \\
        Metallicity [$Fe/H$] & {$-0.09\pm0.08$} & \cite{mann2015} \\    
        Mass $M_\star$ ($M_\odot$) & $0.262 \pm 0.021$ & This work \\
        Radius $R_\star$ ($R_\odot$) & $0.275 \pm 0.009$ & This work \\
        Density $\rho_\star$ ($\mathrm{g~cm^{-3}}$) & $17.7^{+1.6}_{-1.5}$ & This work \\
        Luminosity $L_\star$ ($L_\odot$) & $0.0085 \pm 0.0008$ & This Work \\
        Rotational Velocity $v \sin i$ (km s$^{-1}$) & $2.62 \pm 0.26$ & \citet{mas-buitrago2024} \\
        \hline
        
    \end{tabular}
    \caption{Measured and Derived Stellar Properties for GJ 3378}
    \label{tab:stellar}
\end{table}

\subsection{Archival SPIRou Spectroscopy}

\citet{moutou2024} presented 176 observations\footnote{Moutou et al.~stated that GJ 3378 had been observed 181 times, but the accompanying data included only 176 RVs.} of GJ 3378 from the stabilized, NIR ($\lambda \lambda = 0.95-2.5 \mu$m) spectro-polarimeter/Doppler spectrometer SPIRou \citep{donati2020} on the 3.6\,m Canada-France-Hawaii Telescope on Maunakea in Hawaii.  RVs were extracted from SPIRou spectra using the \texttt{wapiti} \citep{ouldelhkim2023} line-by-line pipeline.

In this analysis, we have removed any SPIRou RVs marked with data quality flags, as well as excluded any observations with signal-to-noise (SNR) values less than half that of the median for the entire time series.  After these cuts, the time series consists of 170 RVs spanning a baseline of 866 days.  The velocities have an RMS scatter of 4.44 \mps and a median uncertainty of 1.46 \mpsp.  The SPIRou data provided by \citet{moutou2024} also include the \dets stellar temperature variability metric \citep{artigau2024}, which we incorporate in our analysis of GJ 3378's magnetic activity.

The full time series of all RVs included in this analysis is shown in Figure \ref{fig:orbit}.  We note here that there are also 23 RVs of GJ 3378 from the HIRES spectrometer presented in \citet{butler2017}.  Given the low number and cadence of these velocities (23 visits over 11 years), we have not included them in our analysis.

\subsection{TESS Photometry}

GJ 3378 has been observed by the Transiting Exoplanet Survey Satellite \citep[TESS;][]{ricker2015} in four Sectors (19, 59, 60, 73) between November 2019 and January 2024.  We downloaded the 2-minute cadence lightcurves as extracted by the Science Processing Operations Center \citep[SPOC;][]{jenkins2016} pipeline from The Mikulski Archive for Space Telescopes \citep[MAST;][]{tess2021} using the \texttt{lightkurve} \citep{lightkurve2018} software package.

Outliers were removed from each sector by removing data beyond 4 standard deviations from the median flux. No additional detrending was performed. The sectors were stitched together for joint analysis, amounting to 57,747 total observations with an average precision of 750 ppm.

\begin{wrapfigure}[36]{r}{0.5\textwidth}
\includegraphics[width=0.5\textwidth]{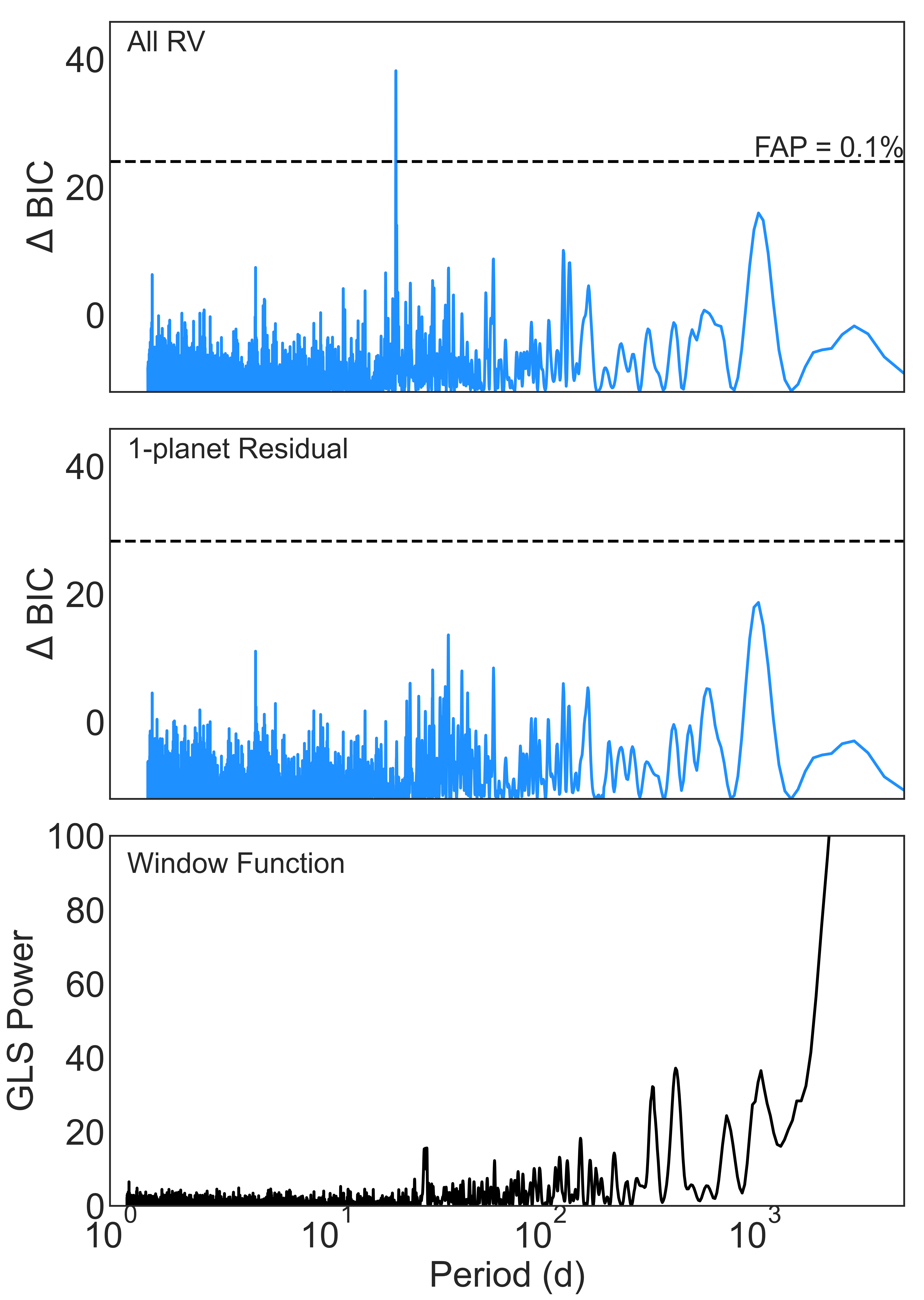}
\caption{RV periodograms for GJ 3378.  \emph{Top:} \texttt{rvsearch} periodogram of all RVs.  \emph{Middle:} \texttt{rvsearch} periodgram of the RVs after removing the 21.5-day planet candidate.  The dashed horizontal lines in both RV periodograms indicate the FAP = 0.1\% threshold as determined by \texttt{rvsearch}.  \emph{Bottom:} window function of the RV sampling, as determined by the GLS algorithm.}
\label{fig:rv_ps}
\end{wrapfigure}

\section{Stellar Properties}

GJ 3378 is a bright, nearby star, which allows for highly precise estimates of its fundamental properties.  It is not known to have any stellar companions \citep{elbadry2021}.  We have combined our high-resolution, high-SNR spectra from HPF and NEID with literature data to constrain the stellar parameters.  In Table \ref{tab:stellar}, we list the measured and derived stellar properties of GJ 3378, as well as the literature references for those estimates.

For the stellar distance $d = 7.730^{+0.002}_{-0.001}$ pc, we adopted the ``photogeometric" distance from \citet{bailerjones2021}, which uses a direction-dependent Galactic extinction model to correct for reddening.  This distance is well within the $1\sigma$ uncertainties of the purely geometric distance estimate. We adopted $T_{eff}=3340\pm60$ K, $\log g_\star=4.97\pm0.06$, and $\mathrm{[Fe/H]}=-0.09\pm0.08$, calculated in \cite{mann2015}, as the spectroscopic parameters. We calculated model-dependent stellar parameters (e.g., $M_\star$, $R_\star$) by fitting the spectral energy distribution (SED) with the \texttt{EXOFASTv2} package \citep{exofast}. \texttt{EXOFASTv2} models the observed broadband photometry using the MIST model grids \citep{Dotter2016,Choi2016}. For the SED fit, we ignored the effects of reddening and applied Gaussian priors on (i) broadband photometry from 2MASS \citep{monet2003} and WISE \citep{wise}, (ii) spectroscopic parameters from \cite{mann2015}, and (iii) parallax measurements from Gaia DR3 \citep{gaiadr3}. \autoref{tab:stellar} contains the derived stellar parameters.

\section{Analysis}
\label{sec:analysis}

\subsection{Frequency Analysis}
\label{sec:periodograms}

We began our analysis of the RV time series by searching for periodic signals in the velocities.  We used the generalized Lomb-Scargle \citep[GLS;][]{zk2009} periodogram, which is a modification of the frequentist Fourier transform for unevenly-sampled data.  We also used the \texttt{rvsearch} \citep{rosenthal2021} periodogram, which at every test frequency performs a fast model comparison between models with and without a Keplerian at that frequency.  We find this periodogram is especially well suited for data sets such as analyzed here, as the model comparison is useful when various instruments have significantly different mean uncertainties, and the zero-point offsets between instruments are allowed to float.

\begin{wrapfigure}[40]{r}{0.5\textwidth}
\includegraphics[width=0.5\textwidth]{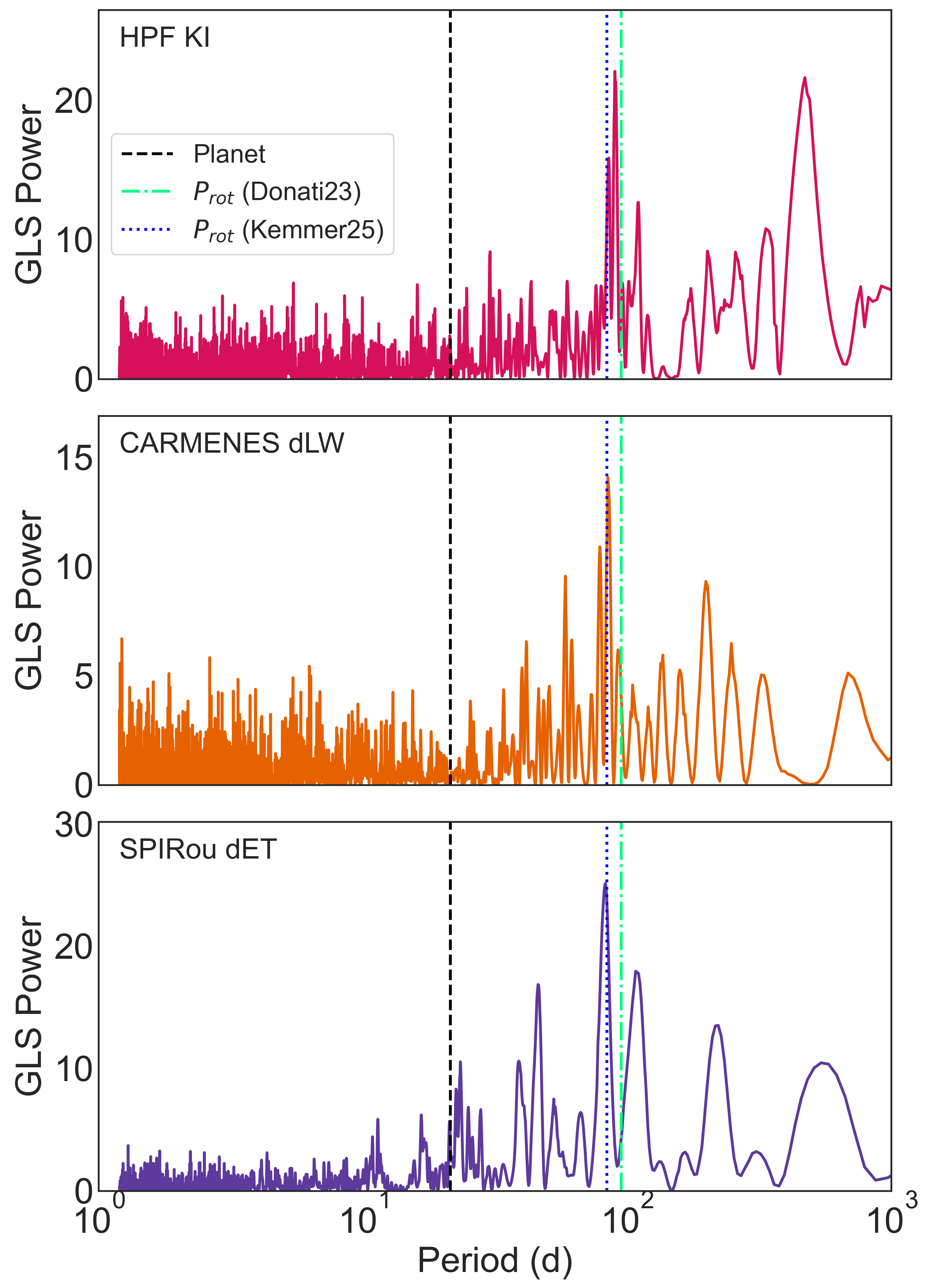}
\caption{GLS periodograms of spectral activity indicators from the spectrometers used in this study.  From top to bottom, we show periodograms of the KI EW from HPF, the CARMENES dLW line shape parameter, and the SPIRou \dets metric.  Vertical lines indicate the period of the planet candidate, as well as stellar rotation periods as determined by \citet{donati2023} and \citet{kemmer2025}.  All instruments/indicators are in broad agreement on the stellar rotation period, and there is no evidence for stellar variability near the period of the planet candidate.}
\label{fig:act_ps}
\end{wrapfigure}

We find consistent results across both periodograms.  In Figure \ref{fig:rv_ps}, we show the \texttt{rvsearch} periodogram of the combined RV series.  It shows a single strong peak at $P = 21.45$d.  The peak well exceeds the FAP = 0.1\% false-alarm threshold as determined by \texttt{rvsearch}, which estimates a peak's significance by comparing to the overall distribution of peak powers throughout the power spectrum.  Upon modeling and removing this signal, we see no additional significant peaks in the periodogram.  Figure \ref{fig:rv_ps} also includes the window function for the combined time series; essentially, the power spectrum of the temporal sampling.  We note that the low-amplitude peak in the window function near the period of the planet candidate lies at the $29.5$d lunar orbital period, and appears to be driven primarily by the SPIRou observing cadence.

In order to ensure the observed periodicity was not being driven by systematics in a single spectrometer, we computed multiple instances of the periodogram, each time leaving out one instrument's RVs.  While the power at the planet period varies depending on which data are excluded, the predominant peak always occurs near 21.5 days.  We do note that when analyzing the SPIRou RVs alone, we recover the 24.7-day peak described by \citet{moutou2024}.  The periodograms of all possible combinations of RV sets are shown in Figure \ref{fig:ps_multi}; we find that the shorter 21.5d period is preferred in all combinations of 3 or more instruments, and in any 2-instrument sets that exclude SPIRou.

\subsection{Stellar Variability}
\label{sec:activity}

When identifying exoplanets in RV time series, it is important to assess the likelihood that the observed signal is in fact driven by stellar surface variability.  Magnetic inhomogeneities such as starspots will alter stellar line profiles through asymmetric flux reduction and suppression of convection \citep[e.g.][]{dumusque2014,siegel2024}, and induce quasiperiodic RV signals at the stellar rotation period and its harmonics/aliases \citep[e.g.][]{robertson2014,lubin2021}.

GJ 3378 is understood to be a slow rotator \citep{mas-buitrago2024}, although its apparent rotation period has been shown to depend on the variability metric used to measure it, and perhaps on the epoch of measurement.  \citet{donati2023}\footnote{We note here that \citet{moutou2024} mistakenly cited a different study by Donati et al.~when referring to this measurement.} found a rotation period of $95.1 \pm 2.3$d via Gaussian Process \citep[GP;][]{ambikasaran2015} regression of fluctuations in the stellar longitudinal magnetic field, as measured by SPIRou spectro-polarimetry.  \citet{moutou2024} conducted periodogram and GP analysis of the SPIRou \dets metric, and found a shorter period of 82.9d.  They explained the difference as potentially being due to the nature of the measurements analyzed.  \dets traces variability due to individual starspots or spot complexes, and will be subject to differential rotation depending on the spot latitude.  On the other hand, the spectro-polarimetry is more closely tracing the global magnetic field.  \citet{kemmer2025} recently estimated a rotation period of $83.82 \pm 1.05$d from periodogram analysis of CARMENES spectroscopic activity indicators, consistent with the shorter period seen in SPIRou \detp.

As has been demonstrated already by previous studies, the rotational modulation of GJ 3378 can be seen broadly consistently across a wide range of activity indicators.  As a concise example, in Figure \ref{fig:act_ps}, we show GLS periodograms of separate spectral activity indicators from each of HPF, CARMENES, and SPIRou.  The NEID time series does not have sufficient time baseline in either the pre- or post-break data separately to achieve sensitivity to the stellar rotation period, so we do not include NEID in Figure \ref{fig:act_ps}.  The CARMENES dLW values show strong power at very long periods, which appears to be systematic in nature; hence, for visual clarity, we show the periodograms out to 1000d.

The HPF KI EW values show multiple periodogram peaks near the previously-published rotation periods from SPIRou and CARMENES.  Currently, the strongest peak occurs at $89.99$d, which is intermediate between the 95d spectro-polarimetric period and the shorter dLW/\dets estimates.  However, if we restrict our analysis to data earlier than September 2024, the highest peak instead lies at 78d.  This demonstrates that---as expected---the rotational modulation is quasiperiodic in nature, and the observed period will depend on the season(s) over which the star is observed.  We will discuss the stellar rotation further in \S\ref{sec:discussion}, but we note here that our HPF data show rotational modulation that is qualitatively consistent with studies by \citet{donati2023}, \citet{moutou2024}, and \citet{kemmer2025}.

None of the activity indicators considered show any significant periodogram power near the period of the candidate planet.  Furthermore, we observe no correlations between RV measurements and the spectroscopic activity indicators.  As described in \S\ref{sec:model}, we attempted to include a GP correlated noise component in our RV model, with broad priors on the hyperparameters loosely informed by our frequency analysis of the activity indicators.  However, this model suffered from over-parameterization, and we conclude that the inclusion of a correlated noise model is not supported by the data.  Therefore, our analysis agrees with that of \citet{moutou2024} in that we find no evidence that stellar magnetic activity currently impacts the RVs at a measurable level, and the frequency space near that of the candidate planet's period appears particularly unbiased by stellar variability.

As an additional check that the observed RV signal does not arise from stellar variability, we examined the evolution of the signal over time.  If the signal is the result of quasiperiodic rotational modulation, we would expect the power to demonstrate significant changes in both period and amplitude as spots appear, dissipate, and reappear with different configurations.  

\begin{wrapfigure}[]{r}{0.6\textwidth}
\includegraphics[width=0.6\textwidth]{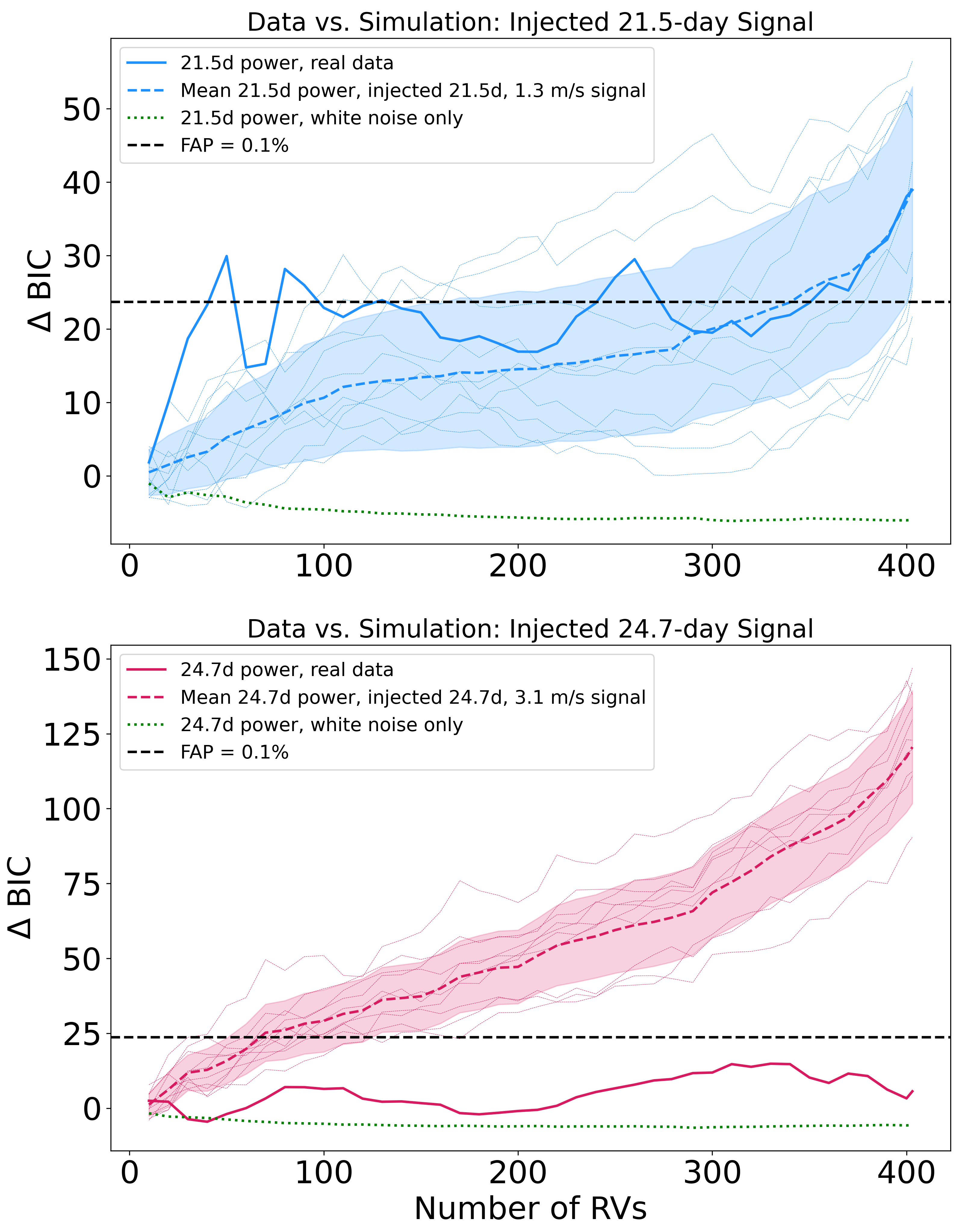}
\caption{\texttt{rvsearch} periodogram power of GJ 3378 RVs, showing the evolution of power as data are added.  The top panel shows power at the 21.5-day period.  The solid curve shows power for the data, and the dashed curve shows the average results for 100 injections of a synthetic planet with orbital parameters from Table \ref{tab:orbit}.  The shaded region shows $1\sigma$ bounds, and individual draws from the test are shown as faint lines.  The mean power at 21.5d for simulations containing only white noise is shown as a dotted green line.  The bottom panel is similar to the top, but for power at the 24.7-day period.  Here, the injected planet matches the description of \citet{moutou2024}.}
\label{fig:stacked_ps}
\end{wrapfigure}

In Figure \ref{fig:stacked_ps}, we show the strength of the strongest peak (within $\pm 0.5$d) near each of the 21.5- and 24.7-day signals from \texttt{rvsearch} periodograms as RVs are added sequentially.  For comparison, we have computed the equivalent periodograms for two synthetic RV time series.  The synthetic RVs retain the timestamps and errorbars of the real data, but we have injected Keplerians matching the orbits of GJ 3378b proposed here and in \citet{moutou2024}.  At each timestamp, we have added white noise, drawn from a Gaussian with width equal to the quadrature sum of the corresponding errorbar and the instrumental jitter value listed in Table \ref{tab:orbit}.  These trials are performed 100 times and averaged, in order to show the mean expected behavior for each scenario.  This test is equivalent to those performed in, e.g., \citet{mortier2017} or \citet{lubin2021}.

We see that for both the real and synthetic data, the power in the 21.5d peak increases as data are added, but not uniformly, particularly at small numbers of RVs.  We believe this is a result of combining data from multiple spectrometers, each of which has different precision and noise properties.  We find that if we add data in order of largest to smallest errorbar size, rather than in time order, the power increases smoothly and monotonically.  We show the evolution of the 21.5d power if data are sorted by errorbar size in Figure \ref{fig:stacked_ps_esort}.  Given this behavior, and the fact that individual draws from our synthetic data tests show similar non-monotonic power increase over time, we conclude that the evolution of the 21.5-day signal is consistent with a coherent Keplerian plus white noise.

On the other hand, a significant peak at the 24.7d period proposed by \citet{moutou2024} never appears in the periodogram of the real data.  Looking at the evolution of the full periodogram as data are added, we see that the power remains concentrated in the 21.5d peak, rather than being distributed among a ``comb" of frequencies as would be expected for a quasiperiodic stellar signal.  The evolution of the full periodogram in the region of the candidate periods is shown in Figure \ref{fig:stacked_ps_time}.  Lastly, we see that if the planet had a 24.7d period and 3.1 m/s Doppler semiamplitude, we should have recovered it at much higher significance, and it would not create a strong periodogram peak at 21.5d.

\subsection{Orbital Model}
\label{sec:model}

We modeled the signal as a Keplerian using the \texttt{RadVel} MCMC analysis package \citep{fulton2018}.  In order to determine the most appropriate solution to the current data, we compared models with and without an exoplanet, as well as including complexities such as an eccentric orbit or a GP correlated noise term.  Based on the outcomes of this modeling, alongside the activity analysis discussed in \S \ref{sec:activity}, we find that the most likely explanation for the observed RV signal is an exoplanet, GJ 3378b.

For all orbital parameters, we used broad, uninformative priors.  In particular, a broad prior on the orbital period was necessary to ensure the model could explore solutions with both the 21.5d period implied by our periodogram analysis and the 24.7d period proposed by \citet{moutou2024}. For most parameters, we used uniform priors, with the exception of the instrumental jitter terms ($\sigma$), which encode additional white noise in an RV series that is in excess of the errorbars.  For those, we used the modified Jeffreys prior suggested by \citet{ford2007} for comparing RV models with different numbers of planets.  For the jitter priors, we adopt a minimum of 0 \mpsp, a maximum of $10.8$ \mps (3 times the RMS of the full RV time series), and a ``knee" value of 1 \mps for all spectrometers, although we note that our results are insensitive to these specific choices.

\begin{table}[]
    \centering
    \begin{tabular}{l l l}
        \hline \hline
        Parameter & Posterior & Prior \\
        \hline
        \\
        \multicolumn{3}{c}{\it Measured Planet Parameters} \\
        \\
        Period (d) & $21.45 \pm 0.01$ & Uniform[0.5, 10000] \\
        Time of conjunction $T_c$ (BJD-TDB) &  $2460292.2 \pm 0.6$ & Uniform[2460284.51, 2460306.01] \\
        Doppler semiamplitude $K$ (\mpsp) & $1.3 \pm 0.2$ & Uniform[0.01, 100] \\
        $\sqrt{e} \cos \omega$ & 0.0 & Fixed \\
        $\sqrt{e} \sin \omega$ & 0.0 & Fixed \\
         & & \\
         \multicolumn{3}{c}{\it Measured Instrument Parameters} \\
          & & \\
          $\gamma_{\textrm{CARMENES}}$ (\mpsp) & $0.2 \pm 0.3$ & Uniform[-20, 20] \\
          $\gamma_{\textrm{HPFpre}}$ (\mpsp) & $0.0 \pm 0.4$ & Uniform[-20, 20] \\
          $\gamma_{\textrm{HPFpost}}$ (\mpsp) & $0.2 \pm 0.3$ & Uniform[-20, 20] \\
          $\gamma_{\textrm{NEIDpre}}$ (\mpsp) & $0.4 \pm 0.3$ & Uniform[-20, 20] \\
          $\gamma_{\textrm{NEIDpost}}$ (\mpsp) & $-0.1 \pm 0.3$ & Uniform[-20, 20] \\
          $\gamma_{\textrm{SPIRou}}$ (\mpsp) & $-0.1 \pm 0.3$ & Uniform[-20, 20] \\
          $\sigma_{\textrm{CARMENES}}$ (\mpsp) & $1.7 \pm 0.2$ & Jeffreys[0, 10.8, 1.0] \\
          $\sigma_{\textrm{HPFpre}}$ (\mpsp) & $2.5 \pm 0.4$ & Jeffreys[0, 10.8, 1.0] \\
          $\sigma_{\textrm{HPFpost}}$ (\mpsp) & $2.0 \pm 0.3$ & Jeffreys[0, 10.8, 1.0] \\
          $\sigma_{\textrm{NEIDpre}}$ (\mpsp) & $0.3 \pm 0.3$ & Jeffreys[0, 10.8, 1.0] \\
          $\sigma_{\textrm{NEIDpost}}$ (\mpsp) & $0.3 \pm 0.3$ & Jeffreys[0, 10.8, 1.0] \\
          $\sigma_{\textrm{SPIRou}}$ (\mpsp) & $4.1 \pm 0.3$ & Jeffreys[0, 10.8, 1.0] \\
           \\
           \multicolumn{3}{c}{\it Derived Planet Parameters}
           \\
           Semimajor axis $a$ (AU) & $0.09673 \pm 0.0008$ & \\
           Minimum mass $m \sin i$ ($M_\oplus$) & $2.3 \pm 0.4$ & \\
           Instellation $S$ ($S_\oplus$) & $0.91 \pm 0.09$ & \\
           Equilibrium temperature $T_{eq}$ (K) & $272 \pm 7$ & \\

            \hline
        
    \end{tabular}
    \caption{Measured and Derived Orbital Parameters for GJ 3378b.  Equilibrium temperature assumes zero albedo.}
    \label{tab:orbit}
\end{table}

All MCMC runs used 5 ensembles of 150 random walkers each, initialized from a Powell maximum-likelihood optimization.  A set of ``burn-in" samples was discarded after the chains reached a Gelman-Rubin \citep[G-R;][]{ford2006} convergence criterion of 1.03, and the chains terminated either after $10^4$ steps per walker, or when the G-R statistic dropped below 1.001.

We find that a model with one planet on a circular orbit is preferred.  The 1-planet model is clearly preferred over a zero-planet model, where only the instrument zero points ($\gamma$) and jitters are allowed to vary.  \citet{kass1995} suggest approximating the Bayes factor (BF) between two models from the Bayesian information criterion (BIC) via $\ln (\textrm{BF}) \sim \frac{\Delta \textrm{BIC}}{2}$, where $\ln (\textrm{BF}) > 5$ is considered significant.  By that metric, we approximate $\Delta$ BIC $\sim 69.4$, corresponding to $\ln (\textrm{BF}) = 34.7$.  The posterior values of our 1-planet circular model are listed in Table \ref{tab:orbit}.  This orbital model is plotted over the time-series RV data in Figure \ref{fig:orbit}, alongside the RVs phase-folded to the model.

\citet{moutou2024} found a modestly significant eccentricity $e = 0.36^{+0.13}_{-0.16}$ for the planet, and we considered models where the eccentricity (parameterized as $\sqrt{e} \sin \omega$ and $\sqrt{e} \cos \omega$) were allowed to vary.  While eccentric models are still greatly preferred over 0-planet models ($\ln (\textrm{BF}) = 28.4$), we found that the circular model ($\ln (\textrm{BF}) = 34.7$) had a higher likelihood despite fewer free parameters, and is thus the best model to the current data.  When we allowed the eccentricity to vary, we found $e = 0.08^{+0.14}_{-0.08}$, consistent with zero to within $1\sigma$.  Furthermore, the planet parameters listed in Table \ref{tab:orbit} do not change significantly regardless of whether the model allows for nonzero eccentricity.

Despite the fact that our periodogram analysis did not show evidence of significant variability other than the 21.5-day exoplanet candidate, we nonetheless attempted models which included a GP correlated noise component.  We used the quasiperiodic GP kernel included in \texttt{radvel}, which has been used in many RV studies to approximate rotationally-modulated stellar variability in time-series data \citep[e.g.][]{haywood2014,polanski2024}.  Our correlated noise model used broad, uninformative priors on all the GP hyperparameters except for the periodic term, for which we adopted a Gaussian prior centered on $P_{rot} = 90 \pm 5$ days, based on our knowledge of the rotation period from activity tracer analysis.  The GP model yielded a lower $\Delta \textrm{BIC}$ improvement than the planet-only model, and was degenerate with the white-noise jitter terms.  As was the case when considering eccentricity, the orbital parameters of planet b were unchanged regardless of whether we included a GP term.  Based on the outcome of the GP models, we conclude that any rotational modulation of the spectra of GJ 3378 is below the current RV detection threshold of the instruments utilized in this study.

\begin{figure}
\includegraphics[width=\textwidth]{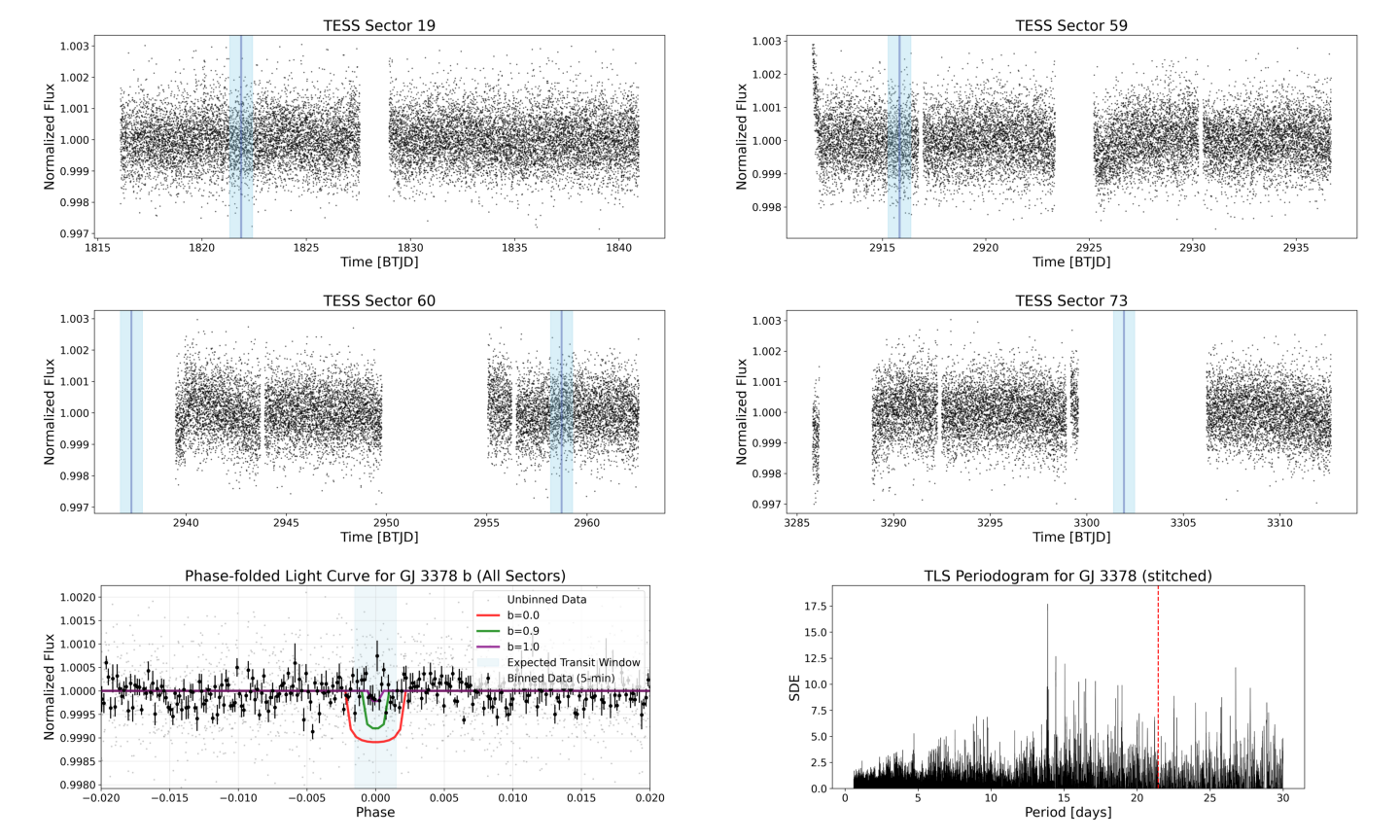}
\caption{TESS photometry of GJ 3378, detailing our search for a transit of planet b.  The top four panels show individual TESS sectors of GJ 3378, while the lower-left panel shows all TESS photometry binned and phase-folded to the period of planet b.  Overplotted on the phase-folded data are transit models of the planet for a range of impact parameters, conservatively assuming a planetary radius $R = 1R_\oplus$.  In the lower right, we show a transit-least-squared (TLS) periodogram of all TESS photometry, with the vertical red line indicating the period of planet b.  We see no evidence for transits of GJ 3378b, or of any additional bodies in the system.}
\label{fig:tess}
\end{figure}

\citet{moutou2024}, when modeling SPIRou and CARMENES RVs jointly, found a 2-planet solution with a second planet at 43d.  We did not consider a 2-planet model in our own analysis, as periodograms of the residuals to our 1-planet model (Figure~\ref{fig:rv_ps}) showed no significant power at any period.

\subsection{Transit Search}

Based on the orbit model shown in Table \ref{tab:orbit}, three of the four TESS Sectors (19, 59, 60) contain a window during which GJ 3378b would have transited, provided its orbit was appropriately inclined.  The geometric transit probability, $R_\star/a \sim 1.4$\% is low, but the scientific value of a transiting, potentially terrestrial-mass exoplanet within 10 pc would be extremely high.

We see no evidence that GJ 3378b is a transiting exoplanet.  In Figure \ref{fig:tess}, we show the four TESS sectors containing transit windows for GJ 3378b, as well as the multi-sector lightcurve phase-folded to the planet's period. As shown in Figure \ref{fig:tess}, even with a conservatively-estimated planet radius of $R = 1 R_\oplus$, we would expect to see a high-amplitude signal in the event of a transit.  To explore the possibility of transits in the lightcurve of GJ 3378 more broadly, we computed the Transit Least Squares \citep[\texttt{TLS};][]{hippke2019} periodogram of the full TESS time series.  The \texttt{TLS} periodogram is included in Figure \ref{fig:tess}.  Adopting the convention that peaks with signal detection efficiency (SDE) greater than 7 are significant, we see no significant peaks at periods less than half the duration of a TESS sector (i.e.~the length of a single TESS orbit).  At longer periods, we see a strong peak at 13.9 days, corresponding to half a TESS sector baseline, and a number of additional marginally significant peaks that we attribute to incomplete sampling at longer periods.  As shown in Figure \ref{fig:tess}, the TESS lightcurve fully covers three transit windows for GJ 3378b, and we do not see a significant peak in the \texttt{TLS} periodogram corresponding to its period.

\section{Discussion}
\label{sec:discussion}

\subsection{Comparison with \citet{moutou2024}}

Our best-fit parameters for the orbit of GJ 3378b are inconsistent with those of \citet{moutou2024} by 47 combined standard deviations in period, and 2.2 combined standard deviations in Doppler semi-amplitude.  The SPIRou-only analysis also suggested a significant eccentricity for the planet, which our model does not prefer.

As detailed in \S\ref{sec:analysis}, the signal seen by SPIRou appears to be inconsistent with RVs from CARMENES, HPF, and NEID, either separately or in combination.  This was partially detailed in \citet{moutou2024}, who analyzed the CARMENES RVs both alone and alongside the SPIRou RVs.  CARMENES alone indicated a 2-planet solution, with eccentric planets at periods of 21d and 43d.  The CARMENES+SPIRou solution included only planet b, and the model had a period of 25d and amplitude $K = 1.2$ \mpsp.  Moutou et al.~adopted the SPIRou-only model for GJ 3378b, arguing that the chromatic nature of activity-induced RV noise made interpretations of models combining optical and NIR RVs complicated.  

Given the agreement between optical RVs from CARMENES and NEID with NIR RVs from HPF, the consistency of the 21-day signal over time, and the low levels of activity-induced RV jitter---particularly near the period of the planet---we adopt parameters for GJ 3378b from a model to all available RVs.  We speculate that the 25-day signal seen in SPIRou likely originates from an undiagnosed source of correlated noise, perhaps amplified by the presence of a genuine planet signal nearby in frequency space.  It is perhaps worth mentioning that the beat frequency of the 21.45d orbital period of the planet, and the 29.53d synodic lunar period is 12.42 days, which is very close to the $P/2$ harmonic of the planet period proposed by Moutou et al.  For the purpose of determining the true properties of GJ 3378b, the combined RV time series prefers the shorter period unambiguously (Figure \ref{fig:ps_multi}), and the uncertainties on orbital parameters are significantly improved in the joint model.

\subsection{Stellar Rotation Period}

Taken at face value, there appears to be significant disagreement over the rotation period of GJ 3378.  Analysis of spectroscopic activity tracers from CARMENES \citep{kemmer2025} and SPIRou \citep{moutou2024} suggest a period near 83 days, while spectro-polarimetry from SPIRou \citep{donati2023} indicates a significantly longer period of 95 days.  Our own results presented here lie between these extremes.

The various rotation period estimates are likely consistent when considering the complicated physics of the star, and the differences in the data and analysis techniques used.  The combination of finite starspot lifetimes and differential rotation can cause the rotational modulation to change period and phase, which frequently creates a ``comb" of peaks in a power spectrum of activity indicators, especially when combining data over many rotations.  This problem may be exacerbated for slow-rotating M dwarfs, for which a single rotation period is a significant fraction of an observing season; sampling gaps further confuse the true signal modulation, moving signal power to alias frequencies \citep[e.g.][]{lubin2021}.  Indeed, our activity periodograms (Figure \ref{fig:act_ps}) show complicated structure near the rotation period.  The HPF periodogram, in particular, shows a number of strong peaks, including near the 83-day period identified by \citet{moutou2024} and \citet{kemmer2025}.

As a brief aside, we note that the topic of differential rotation is poorly understood for fully-convective M stars.  Evidence of differential rotation has been observed in mid-to-late M stars \citep[e.g.][]{barnes2005,morin2008,reinhold2013,davenport2015}, but there is no clear expectation for how differential rotation on an M dwarf will compare to the Solar example.  \citet{barnes2005}, using Doppler imaging of 10 G2-M2 dwarfs, found evidence for a decrease in differential rotation with decreasing stellar temperature.  On the other hand, \citet{reinhold2013} found that differential rotational shear \emph{increased} with decreasing stellar temperature in a study of Kepler lightcurves.  While the additional complications related to sampling and activity tracer selection prevent us from definitively claiming evidence of differential rotation for GJ 3378, prior empirical constraints of differential rotation for M stars suggest it can reasonably be considered a contributor to the uncertainty regarding the rotation period.

It is also possible that the various activity metrics considered trace different aspects of the star's activity, and are thus differently sensitive to factors such as spot lifetimes and differential rotation.  \citet{donati2023} argue that the SPIRou spectro-polarimetry traces changes in the longitudinal magnetic field.  Similarly, there is evidence that the HPF KI index is sensitive to magnetic field changes via the Zeeman effect \citep{terrien2022}.  These inhomogeneous data products are furthermore analyzed with different statistical methods (power spectrum analysis, GP regression, etc.), which will not agree on the exact period even with identical inputs.

Ultimately, it is challenging to describe the rotation of a star with a single number.  The available data for GJ 3378 likely trace multiple aspects of the (potentially differentially rotating) star's magnetic field and chromosphere.  These various modulations are subject to changes in period, amplitude, and phase over the $\sim10$-year ($>30$ rotations) baseline of the observations.  Considering these complications, we conclude that the available estimates of the stellar rotation period are more consistent than their statistical uncertainties would suggest, with each offering an incomplete glimpse of the star's true behavior.

\subsection{Potential Habitability and Followup Characterization}

We find a minimum mass of $m \sin i = 2.3 \pm 0.4 M_\oplus$ for GJ 3378b.  This minimum mass is comfortably within the range of planet masses expected to have terrestrial compositions, according to most mass-radius relationships for transiting exoplanets \citep[e.g.][]{weiss2014,chen2017,polanski2024,parc2024}.  Furthermore, the planet has a temperate equilibrium temperature of $T_{eq} = 272 \pm 7$ K, and orbits near the inner edge of the conservative circumstellar liquid-water HZ, according to \citet{kopparapu2013}.  GJ 3378b is therefore among the most potentially Earthlike exoplanets known within the 10-parsec Solar neighborhood.

The canonical definition of the HZ \citep{kasting1993} assumes the planet has an Earthlike atmosphere; whether an HZ exoplanet actually supports surface water depends on the pressure and greenhouse warming provided by its atmosphere.  Given the proximity of M dwarf HZs to the host stars, it is unclear whether terrestrial-mass exoplanets in these orbits can retain atmospheres.  Such planets will be exposed to extreme XUV fluxes, which can strip their atmospheres, particularly when the stars are young, active, and frequently flaring \citep[e.g.][]{ribas2016,dong2017,zahnle2017}.

In order to assess whether it is feasible that GJ 3378b has an extant atmosphere, we estimated its atmosphere retention metric (ARM) as defined by \citet{pass2025}.  The ARM compares $I_{XUV}$, the XUV flux received by an exoplanet over its history, to its escape velocity $v_{esc}$ in order to evaluate the comparative likelihood that a planet is sufficiently massive to preserve its atmosphere in spite of the XUV exposure.  \citet{pass2025} refined estimates of $I_{XUV}$ for mid-to-late M stars, accounting for their extended early active phases, and energetic flares.

Using the \citet{pass2025} relations, our mass and luminosity estimates for GJ 3378 suggest that at the semimajor axis of planet b, the integrated XUV flux is $I_{XUV} = 9.5 \times 10^{19}$ erg~cm$^{-2}$, ie. 48 times that of Earth.  Determining the planetary escape velocity is difficult with only a minimum mass measurement, but if we assume a nearly edge-on orbit (that is, $m \sim m \sin i$), the mass-radius relationship of \citet{parc2024} predicts a radius $R = 1.29 \pm 0.06 R_\oplus$.  This value agrees with other mass-radius relationships \citep[e.g.][]{weiss2014,chen2017}, and with Earth's bulk density.  The minimum mass and associated predicted radius yields an escape velocity $v_{esc} \sim 15$ km s$^{-1}$.  

Our adopted values of $I_{XUV}$ and $v_{esc}$ yield ARM $= -0.13$, placing it directly on the ``cosmic shoreline" \citep{zahnle2017} dividing planets that should (ARM $< 0$), or should not (ARM $> 0$), have lost their atmospheres due to radiative stripping.  The ARM is an interesting order-of-magnitude estimate, but the exact nature of any planet's atmosphere is keenly sensitive to its true mass and radius, and to its exact XUV radiation history.  Finding ARM $\sim 0$ for GJ 3378b essentially indicates that the presence of an atmosphere on this world cannot be ruled out based on simple scaling relationships.

\begin{figure}
    \centering
    \includegraphics[width=\textwidth]{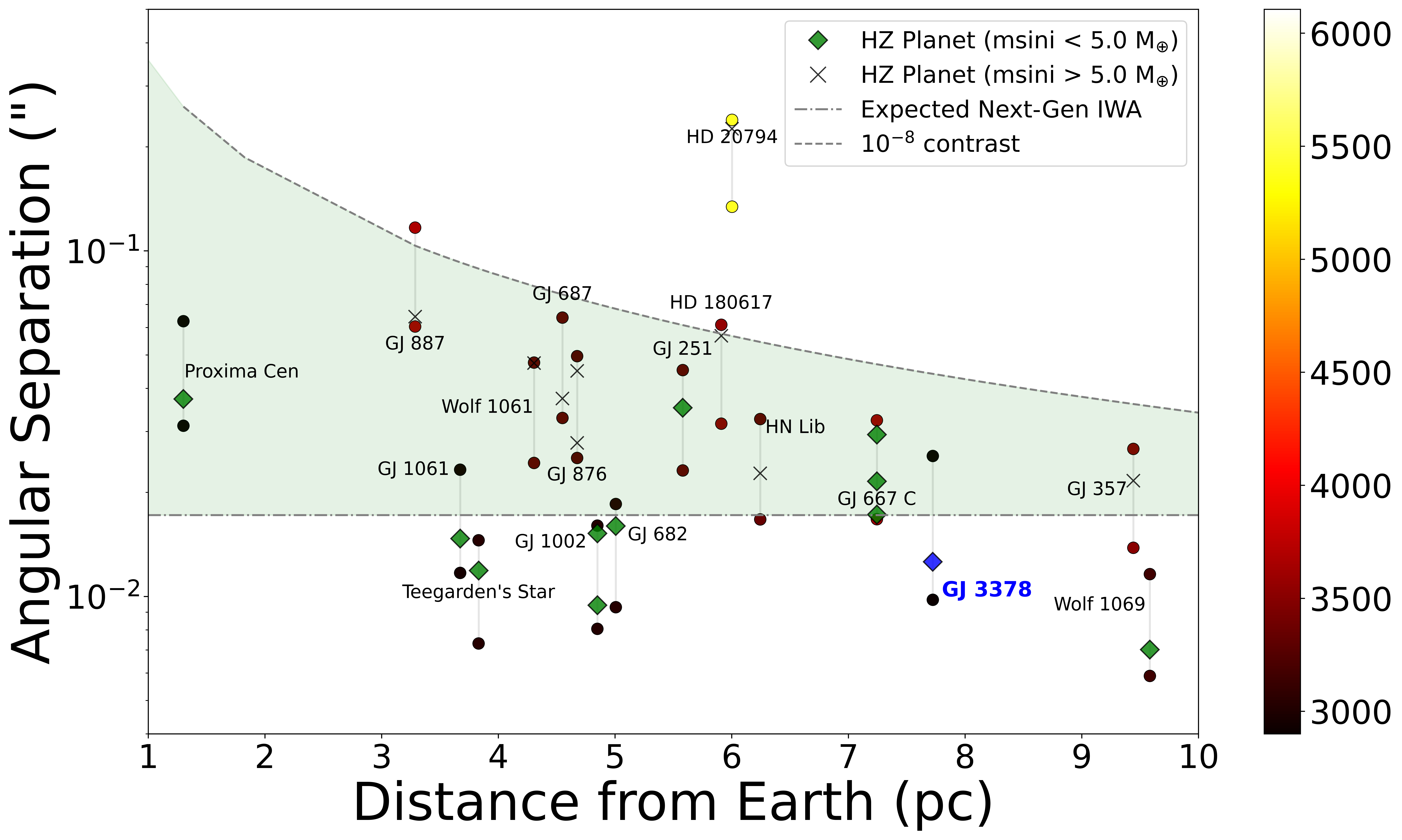}
    \caption{Angular extents of the optimistic HZ \citep{kopparapu2013} for nearby stars hosting exoplanets.  The inner and outer HZ limits are marked with circles colored by stellar effective temperature (in K), and known exoplanets are indicated as diamonds at their projected angular separations.  The green shaded region indicates where planets might be accessible to imaging facilities on 30\,m-class telescopes, limited by inner working angle (IWA) and $10^{-8}$-level contrast.  The HZ of GJ 3378 is partially accessible to future imagers, but imaging GJ 3378b will require them to achieve better IWAs than $2\lambda/D$.}
    \label{fig:imaging}
\end{figure}

As an M dwarf within 10 pc, GJ 3378 is potentially amenable to direct imaging of its HZ with 30\,m-class telescopes.  In Figure \ref{fig:imaging} we show the angular extents of the HZs of nearby stars hosting known exoplanets.  Following \citet{beard2025}, we have estimated the separations at which the planet-to-star contrast ratio drops below $10^{-8}$ for an exoplanet with $R = 2 R_\oplus$ and albedo $A = 0.5$.  The region between this curve and the $2 \lambda/D$ inner working angle (IWA) of a 30\,m telescope provides a very coarse comparative estimate of where HZ exoplanets may be imaged by extremely large telescopes.

The outer region of GJ 3378's HZ is well suited for imaging with 30\,m-class telescopes.  However, at the semimajor axis of planet b, a hypothetical imaging system would have to reach IWAs significantly better than $2 \lambda/D$ to detect it.  It is nonetheless important to characterize the entire planet inventory of the HZ, as the presence of planet b may have significant implications for the dynamical stability or formation history of any additional planets present in the outer HZ.

\section{Conclusion}
We performed a comprehensive analysis of all available RV data for the nearby M dwarf GJ 3378. The addition of our HPF and NEID RVs have resulted in a significant planetary signal with a period of $21.45 \pm 0.01$d and minimum mass $m \sin i = 2.3 \pm 0.4 M_\oplus$.  The planet orbits within the conservative HZ, and a preliminary estimate of its cumulative XUV exposure suggests it may retain an atmosphere.  As an HZ exoplanet within 10 pc, GJ 3378b is potentially amenable to followup characterization with next-generation facilities.

\begin{acknowledgments}

We thank the anonymous referee, who provided insightful comments which improved this manuscript.

The Hobby-Eberly Telescope (HET) is a joint project of the University of Texas at Austin, the Pennsylvania State University, Ludwig-Maximilians-Universität München, and Georg-August-Universität Göttingen. The HET is named in honor of its principal benefactors, William P. Hobby and Robert E. Eberly.  We thank the HET Resident Astronomers for their skillful execution of our HET observations.

We acknowledge the Texas Advanced Computing Center (TACC) at The University of Texas at Austin for providing high performance computing, visualization, and storage resources that have contributed to the results reported within this paper.

These results are based on observations obtained with the Habitable-zone Planet Finder Spectrograph on the HET. The HPF team was supported by NSF grants AST-1006676, AST-1126413, AST-1310885, AST-1517592, AST-1310875, AST-1910954, AST-1907622, AST-1909506, AST-2108493, AST-2108512, AST-2108569, AST-2108801, ATI-2009889, ATI-2009982, ATI-2009554, and the NASA Astrobiology Institute (NNA09DA76A) in the pursuit of precision radial velocities in the NIR. The HPF team was also supported by the Heising-Simons Foundation via grant 2017-0494.

This work was partially completed as part of NASA’s CHAMPs team, supported by NASA, United States under Grant No. 80NSSC21K0905 issued through the Interdisciplinary Consortia for Astrobiology Research (ICAR) program.

This work was performed in part for the Jet Propulsion Laboratory, California Institute of Technology, sponsored by the United States Government under the Prime Contract 80NM0018D0004 between Caltech and NASA.

CIC acknowledges support from NASA under award number 80GSFC24M0006.

Based in part on observations at Kitt Peak National Observatory, NSF’s NOIRLab (Prop. ID 2025A-387657; PI: P.~Robertson), managed by the Association of Universities for Research in Astronomy (AURA) under a cooperative agreement with the National Science Foundation. The authors are honored to be permitted to conduct astronomical research on Iolkam Du\'ag (Kitt Peak), a mountain with particular significance to the Tohono O'odham.  Data presented herein were obtained at the WIYN Observatory from telescope time allocated to NN-EXPLORE through the scientific partnership of the National Aeronautics and Space Administration, the National Science Foundation, and the National Optical Astronomy Observatory.  We thank the NEID Queue Observers and WIYN Observing Associates for their skillful execution of our NEID observations.

This research has made use of the SIMBAD database, operated at CDS, Strasbourg, France, and  NASA's Astrophysics Data System Bibliographic Services.

This work has made use of data from the European Space Agency (ESA) mission
{\it Gaia} (\url{https://www.cosmos.esa.int/gaia}), processed by the {\it Gaia}
Data Processing and Analysis Consortium (DPAC,
\url{https://www.cosmos.esa.int/web/gaia/dpac/consortium}). Funding for the DPAC
has been provided by national institutions, in particular the institutions
participating in the {\it Gaia} Multilateral Agreement.

This paper includes data collected by the TESS mission. Funding for the TESS mission is provided by the NASA's Science Mission Directorate.

The Center for Exoplanets and Habitable Worlds is supported by Penn State and its Eberly College of Science. 

\end{acknowledgments}

\facilities{HET (HPF), WIYN (NEID), TESS, Gaia}

\software{
\texttt{astropy} \citep{astropy18},
\texttt{barycorrpy} \citep{kanodia18b},
\texttt{batman} \citep{kreidberg2015},
\texttt{HxRGproc} \citep{ninan2018},
\texttt{lightkurve} \citep{lightkurve2018},
\texttt{matplotlib} \citep{Hunter07},
\texttt{numpy} \citep{harris20},
\texttt{pandas} \citep{reback2020pandas,mckinney-proc-scipy-2010},
\texttt{RadVel}\citep{fulton2018},
\texttt{scipy} \citep{2020SciPy-NMeth},
\texttt{SERVAL} \citep{zechmeister2018},
\texttt{TLS} \citep{hippke2019}
}

\bibliography{gj3378}{}
\bibliographystyle{aasjournalv7}

\appendix

\begin{figure}
    \centering
    \includegraphics[width=\textwidth]{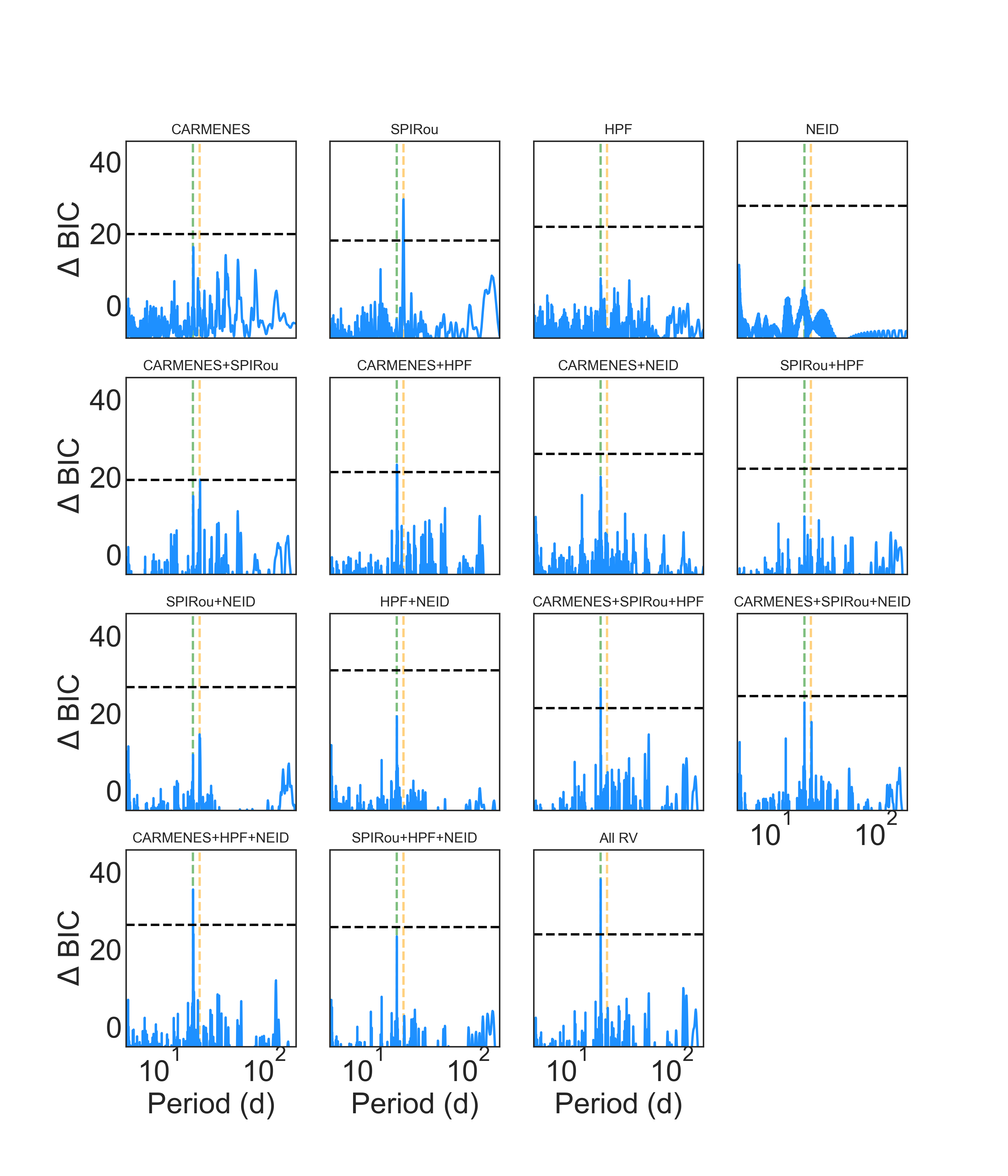}
    \caption{\texttt{rvsearch} periodograms of GJ 3378, using all combinations of spectrometers.  Dashed horizontal lines indicate the FAP = 0.1\% threshold as determined by \texttt{rvsearch}.  Vertical lines show the 24.7-day period for planet c proposed by \citet{moutou2024}, and the 21.5-day period found by our analysis.  The shorter period is preferred by all sets of at least 3 instruments, and by combinations of 2 instruments that do not include SPIRou.}
    \label{fig:ps_multi}
\end{figure}

\begin{figure}
\includegraphics[width=\textwidth]{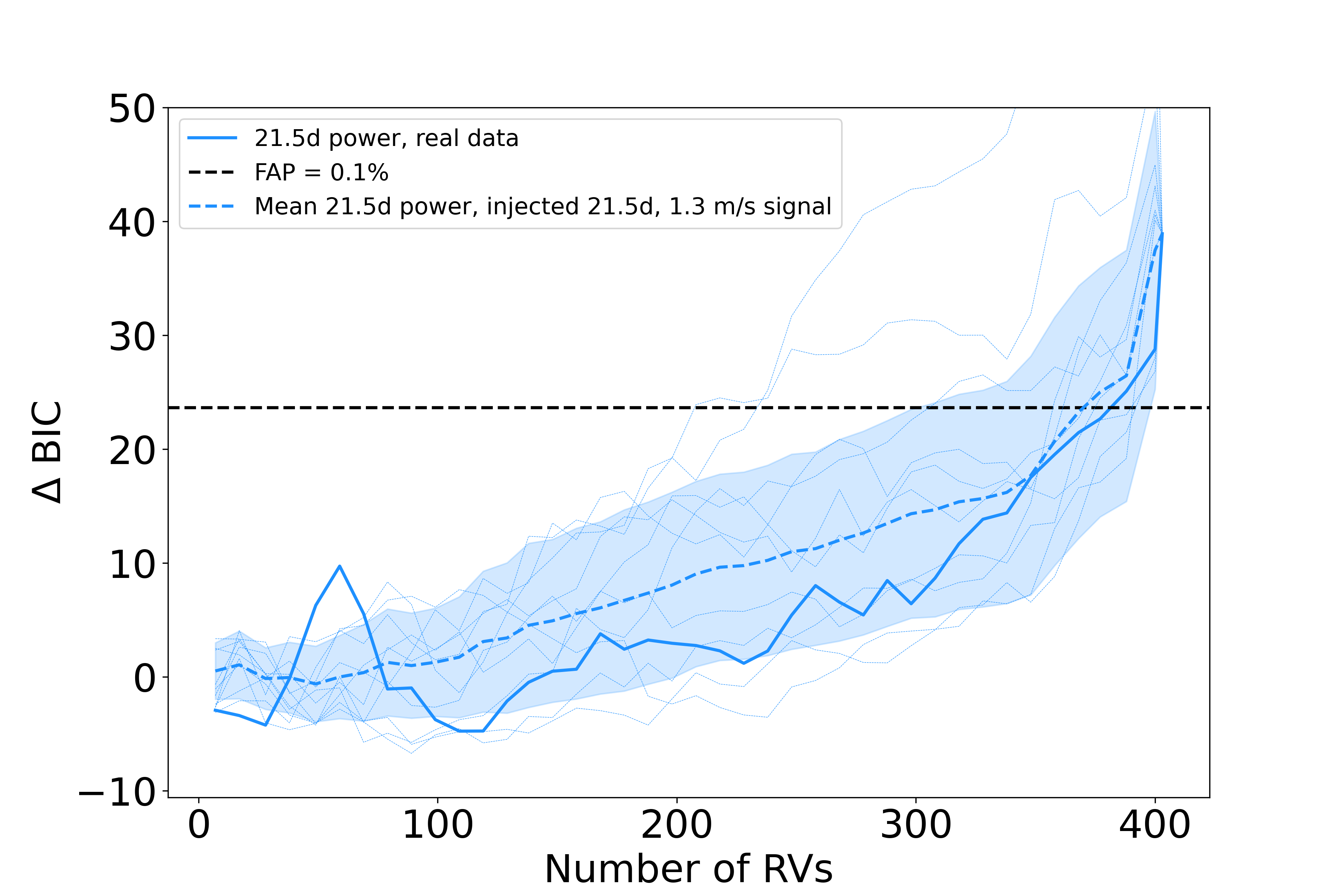}
\caption{Stacked \texttt{rvsearch} periodogram power of GJ 3378 RVs at the period of the 21.5d signal, showing the evolution of power as data are added.  Here, the data are added in order of largest to smallest errorbar size, rather than in time order.  As in Figure \ref{fig:stacked_ps}, we also show the average power increase for 100 trials of a synthetic planet plus white noise (dashed line), the $1\sigma$ bounds (shade), and example draws from the 100 trials (faint lines).  These simulations were also performed with data added in order of decreasing errorbar size.  The power increases steadily as RV information is added.  This behavior suggests that the non-monotonic power increase seen in Figure \ref{fig:stacked_ps} is mostly due to the non-homogeneous noise properties of the RV series.}
\label{fig:stacked_ps_esort}
\end{figure}

\begin{figure}
\includegraphics[width=\textwidth]{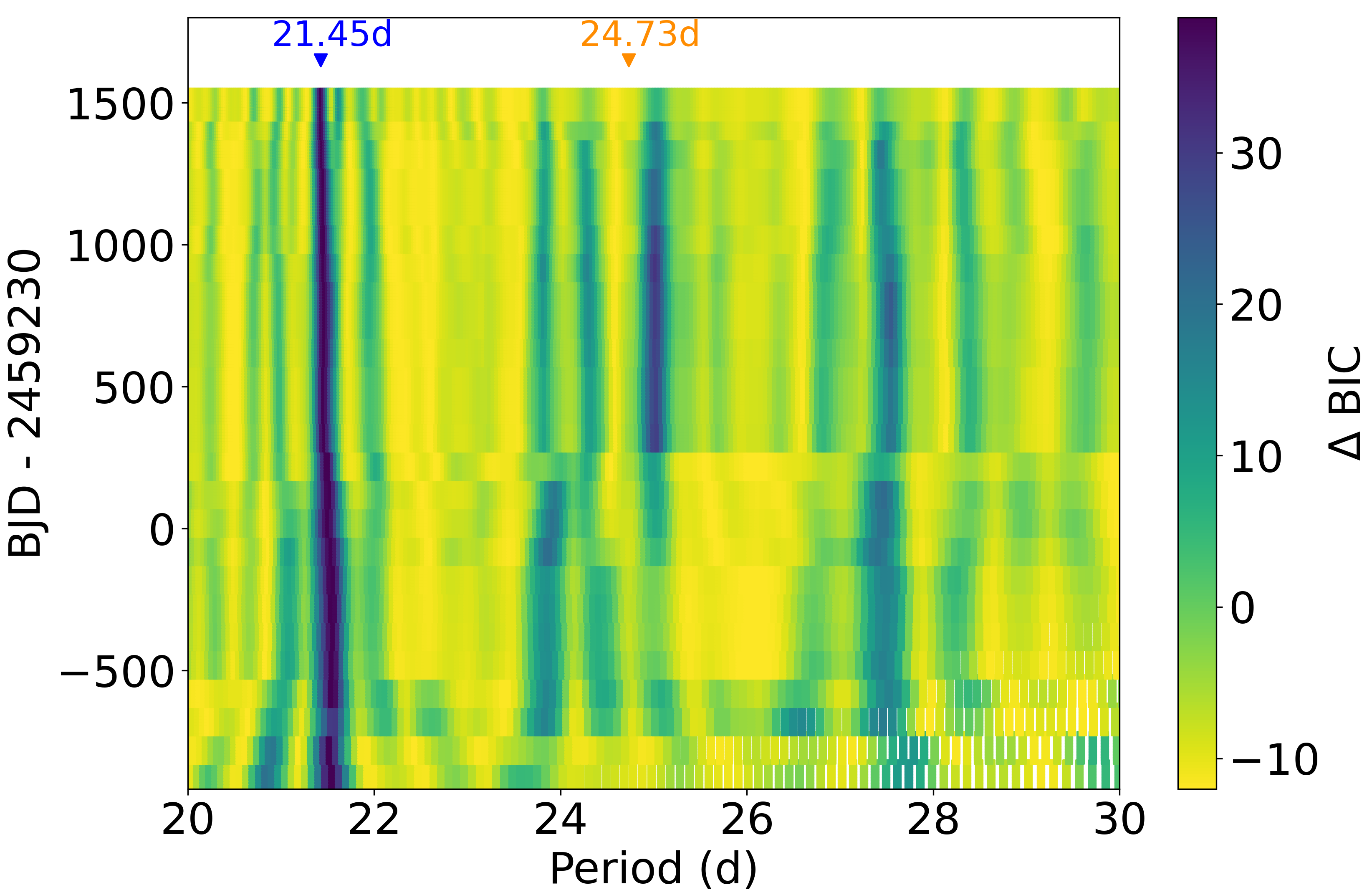}
\caption{Stacked \texttt{rvsearch} periodogram of all GJ 3378 RVs, showing the evolution of the power spectrum as data are added.  We show here the period space containing the exoplanet signal.  The 21.5-day period remains consistent over the baseline of the data and increases in signal-to-noise.  The 24.7-day signal does not appear distinct in the stacked periodogram of all RVs.}
\label{fig:stacked_ps_time}
\end{figure}

\end{document}